\documentclass[]{aa}
\usepackage{graphicx}
\usepackage{txfonts}
\headnote{Letter to the Editor}
\begin{document}
   \title{The cool atmospheres of the binary brown dwarf $\varepsilon$\,Indi B
   \thanks{Based on observations collected with the ESO VLT, Paranal, Chile,
    program 60.A-9245(A).}}

   \author{M. F. Sterzik\inst{1},
          E. Pantin \inst{1,2},
          M. Hartung\inst{1}, N. Huelamo\inst{1}, H.U.
K\"aufl \inst{3}, A. Kaufer\inst{1}, C. Melo\inst{1}, D.
N\"urnberger\inst{1}, R. Siebenmorgen \inst{3}, A. Smette\inst{1} }

   \offprints{M.F. Sterzik, \email{msterzik@eso.org}}

   \institute{European Southern Observatory, Casilla 19001, Santiago 19, Chile
             \and
             DSM/DAPNIA/Service d'Astrophysique, CEA/Saclay, F-91191 Gif-sur-Yvette, France
             \and
             European Southern Observatory, D-85748 Garching b. M\"unchen, Germany}

   \date{Received ; accepted }

   \abstract{We  have imaged $\varepsilon$\,Indi B,
   the closest brown dwarf binary known, with VISIR at the VLT in
   three narrow-band mid-infrared bandpasses located around
   8.6$\mu$m, 10.5$\mu$m and 11.3$\mu$m. We are able to spatially
   resolve both components, and determine accurate mid-infrared photometry
   for both components independently. In particular,
   our VISIR observations probe the NH$_3$
   feature in the atmospheres of the cooler and warmer brown dwarfs.
   For the first time, we can disentangle
   the contributions of the two components, and find that
   $\varepsilon$\,IndiBb is in good
   agreement with recent ``cloud-free'' atmosphere models having
   an effective temperature of $T_\mathrm{eff}=800$\,K.
   With an assumed age of 1 Gyr for the $\varepsilon$\,Indi system,
   component Ba agrees more with $T_\mathrm{eff} \approx 1100$\,K
   rather than with  $T_\mathrm{eff}=1200$\,K, as
   suggested by SPITZER spectroscopic
   observations of the combined $\varepsilon$\,Indi B system (Roellig
   et al., 2004). Even higher effective temperatures appear inconsistent
   with our absolute photometry, as they would imply an unphysical
   small size of the brown dwarf $\varepsilon$\,IndiBa.

   \keywords{stars: low-mass, brown dwarfs --- stars: binaries: general}}
  \authorrunning{Michael F. Sterzik et al.}
  \titlerunning{The cool atmospheres of $\varepsilon$\,Indi B}
   \maketitle
%

\section{Introduction}

$\varepsilon$\,Indi B, the closest brown dwarf known (Scholz et al.,
2003), has been recently discovered as a close binary, consisting of
two brown dwarf components separated by  0.73\arcsec (McCaughrean et
al., 2004). The system has a well-established distance (3.626 pc,
ESA 1997) and age (range 0.8-2Gyr; Lachaume et al., 1999),
and will allow the determination of fundamental physical parameters
like its mass, luminosity, effective temperature, and surface
gravity with unprecedented precision. Extending the analysis towards
the mid-infrared (MIR) offers the opportunity to benchmark
evolutionary and atmospheric models for very low temperatures.
\newline
Adaptive optics assisted near-infrared (H-band) low-resolution
($R\sim1000$) spectroscopy (McCaughrean et al., 2004) of both
components individually lead to a most likely spectral
classification of T1 for $\varepsilon$\,Indi Ba and T6 for
$\varepsilon$\,Indi Bb based on the Burgasser et al. (2002) H$_2$O
and CH$_4$ spectral indices.
Effective temperatures  between 1238K and 1312K were
derived for $\varepsilon$\,Indi Ba and between 835K and 875K for
$\varepsilon$\,Indi Bb, bracketed by assuming the most likely ages
between 0.8 and 2~Gyrs.
However, the comparison of high-resolution
($R\sim50000$) near-infrared spectra of $\varepsilon$\,Indi Ba
(Smith et al., 2003) with synthetic atmosphere spectra of Tsuji
(2002) leads to a much higher $T_\mathrm{eff}=1500 \pm 100$\,K. The
reasons for the large discrepancy are not known, but may be related
to the radius of the brown dwarf or to uncertainties in the
bolometric corrections assumed.
\newline
$\varepsilon$\,Indi B has recently been observed by the SPITZER
Space Telescope in the mid-infrared (Roellig et al., 2004). Their
low-resolution IRS spectrum is a composite spectrum of both
components, as the limited angular resolution of SPITZER does not
allow to resolve it spatially. Roellig et al. claim that their
observation is the first evidence for NH$_3$ absorption in very cool
brown dwarf atmospheres between 10$\mu$m and 11$\mu$m, although they
cannot disentangle the contributions of both components
individually. The SPITZER spectrum matches well a composite model
described by Saumon, Marley \& Lodders (2003, hereafter SML),
assuming cloudless synthetic spectra.
\newline
Here, we report spatially resolved MIR photometry of
$\varepsilon$\,Indi B obtained with VISIR, the new mid-infrared
camera and spectrometer at the VLT (Lagage et al., 2004) during
science verification. Our goal is to constrain the most pertinent
brown dwarf model atmospheres with our data.


\section{Observations}

\begin{figure*}
\centering
\includegraphics[width=6.5cm,angle=90,bb=54 54 300 738]{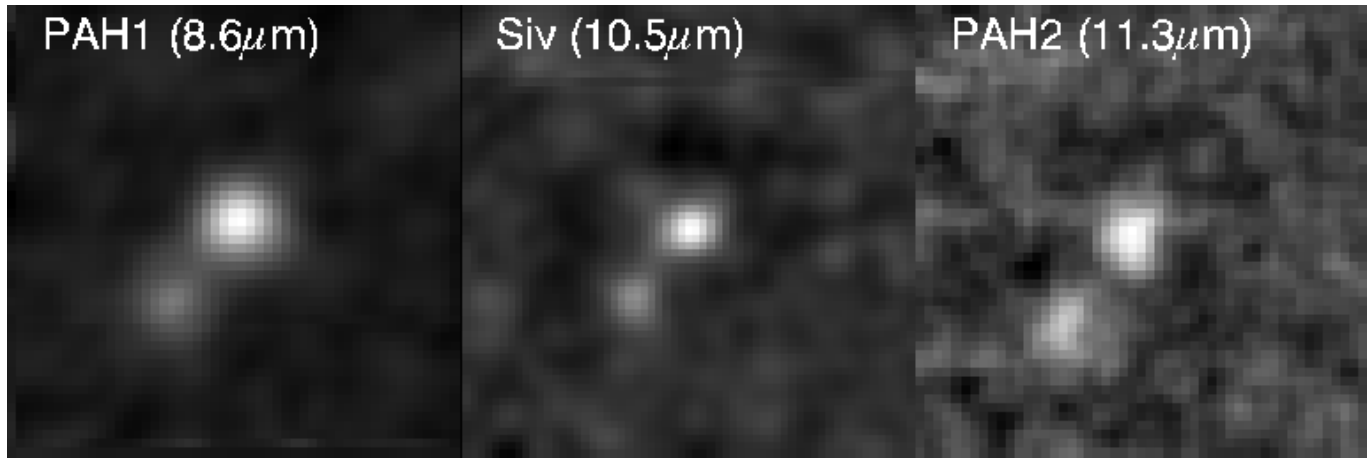}
\caption{VISIR images of $\varepsilon$\,Indi B in three filters
PAH1, S{\sc{iv}} and PAH2. Components a+b are clearly resolved in
all three bandpasses. North is up, and east left:
$\varepsilon$\,Indi Bb is the fainter source to the south-east.}
\label{fig1}

\end{figure*}

$\varepsilon$\,Indi B has been imaged with VISIR mounted on the UT3
(Melipal) of the VLT on Sept. 28, 2004 (in filters PAH1 and PAH2)
and on Sept. 30, 2004 (in filter S{\sc{iv}}) under clear and stable
atmospheric conditions. A nominal pixel scale of 0.075\arcsec was
used in all bands, and standard chopping and nodding techniques
(10\arcsec amplitudes) were employed.
Fig.~\ref{fig1} show the final, co-added and median cleaned images
that were used for sub-sequent photometry. The binary is clearly
resolved, the measured image quality (FWHM) on the final frames is
around $0.4-0.5 \arcsec$ for each point source, slightly above the
diffraction limit.
While the position
angle from our three measurements ($137^\circ\pm6^\circ$) is
consistent with that determined by McCaughrean et al. (2004) using
AO one year earlier ($136.81^\circ\pm0.14^\circ$), our separation
($0.92\arcsec\pm0.06 \arcsec$) is larger ($0.732\arcsec\pm0.002
\arcsec$), and suggests orbital motion.
\newline
A summary of the observing log, filter central wavelengths, half
band-width and total on-source integration times, is given in
Table~\ref{obslog}.
Source count-rates were determined with standard aperture
photometry. Curve-of-growth methods were applied to find the
aperture radii that maximized the signal-to-noise ratio of the
extracted source. Those radii were then also used for the
calibration source to determine the count-rate to flux conversion
factor. For faint sources, the optimal extraction aperture radius is
not always well determined, and the presence of a second companion
does not allow to grow the aperture to more then half of their
distance. The variation of the count-rate to flux conversion factors
with aperture radius was screened for aperture radii of 4, 5 and 6
pixel (corresponding to radii of $0.3\arcsec, 0.375\arcsec,
0.45\arcsec$), and constitutes the main error in the {\sl absolute
flux calibration} of the targets. Different calibrator
stars\footnote{taken from the list of Cohen et al. (1999)} observed
at different airmasses give fully consistent results, and their
contribution to the absolute photometric errors is negligible. {\sl
Flux ratios} between the components, and between different filter
passbands can be obtained with a much higher accuracy, as the errors
are only related the spread of the count rates in the four different
beams, and rather independent on the actual choice of the aperture
radius. Fluxes and ratios together with error estimates are
summarized in Table~\ref{results}.
\newline
We note that the {\sl combined} fluxes of components Ba and Bb agree
with the spectrophotometric fluxes for the $\varepsilon$\,Indi B
system deduced from the SPITZER spectrum shown in Fig.2 of Roellig
et al. (2004), albeit our fluxes are on the low side of their
allowed 25\% error range in absolute flux calibration. Component Ba
dominates the mid-infrared emission, and component Bb contributes
about 1/3 to the total flux density. Our S{\sc{iv}} filter located
at a central wavelength of 10.5$\mu$m is particular well suited to
probe the potential presence of NH$_3$ absorption features in brown
dwarf atmospheres. In fact, a significant change of the relative
spectral shape between both components becomes evident in the
reddest passband around 11.3$\mu$m (PAH2), where the flux of Bb
comes close to that of Ba. The flux ratio S{\sc{iv}}/PAH2 is small
for component Bb, indicating the presence of a strong NH$_3$
absorption band. We find no indication for strong absorption in
component Ba, though.

\begin{table}[ht]
      \caption[]{Log of the VISIR observations of $\varepsilon$\,Indi B.}
         \label{obslog}

         \begin{tabular}{lllllll}
            \hline
            \hline
            \noalign{\smallskip}
            RJD  &   Filter & $\lambda_{cen}$ & $\Delta \lambda$ & AM & Int[s]  &  Calibrator$^\mathrm{a}$ \\
            \noalign{\smallskip}
            \hline
            \noalign{\smallskip}
            53277.6 &  PAH1 & 8.59 & 0.42 & 1.18 & 3638 &
            HD224630\\ & & & & & &HD10550 \\
            53277.7 &  PAH2 & 11.25 & 0.59 & 1.29 & 3587 &
            HD224630 \\
            53279.6 &  S{\sc{iv}} & 10.49 & 0.16 & 1.18 & 3564 &
            HD178345 \\
            \noalign{\smallskip}
            \hline
         \end{tabular}
   \end{table}

\begin{table}[ht]
      \caption[]{Mid-infrared flux densities and flux ratios
      of $\varepsilon$\,Indi B.}
         \label{results}

         \begin{tabular}{lccccc}
            \hline
            \hline
            \noalign{\smallskip}
            Object  &  PAH1 & S{\sc{iv}} & PAH2 & $\mathrm{S{\sc{iv}}\over PAH1}$& $\mathrm{S{\sc{iv}}\over PAH2}$\\
             &[mJy] & [mJy] & [mJy]  & & \\
            \noalign{\smallskip}
            \hline
            \noalign{\smallskip}
            Ba &  $7.4\pm.4$ & $7.2\pm.6$ & $6.8\pm.8$ & $.97\pm.02$& $1.06\pm.03$\\
            Bb &  $3.5\pm.4$ & $3.5\pm.8$ & $5.7\pm1.$ & $1.0\pm.04$& $0.61\pm.08$\\
           \noalign{\smallskip}
            \hline
         \end{tabular}
   \end{table}

\section{Comparison with Atmosphere Models}

\begin{figure}[t]
\centering
\includegraphics[width=7cm, angle=90,bb=54 54 558 668]{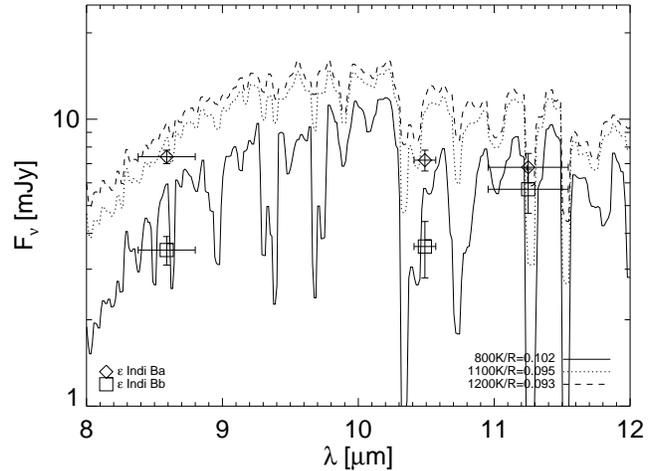}
\caption{Comparison of the VISIR photometry with various atmosphere
spectra from BSL and Burrows (private communication). Radii assumed
are consistent  with evolutionary tracks of Burrows et al. (1997).}
\label{fig2}
\end{figure}

Atmosphere models are typically calculated on a grid of
$T_\mathrm{eff}/g$ pairs, for specific metallicities. Taylor (2003)
lists [Fe/H]=0.056$\pm$0.038 for $\varepsilon$\,Indi A; therefore we
refer to solar metallicities also for $\varepsilon$\,Indi B.
\newline
The object radius $R$ is the main parameter that determines the
absolute spectral flux calibration of the model atmospheres, and is
obtained from evolutionary calculations. Comparing the evolutionary
models of Baraffe et al. (2003) with those of Burrows et al. (1997)
(in the given grid of ages and temperatures), we find that their
radii agree within 2\%. In the following we refer to the Burrows
models, only, and note that different models do not contribute
significantly to uncertainties in the absolute model fluxes.  In
Table~\ref{parameters} we summarize the (sub-)stellar parameters
log\,$g$, $L_\mathrm{bol}$ and $R$ assuming an age of 1Gyr for the
$\varepsilon$\,Indi B system. We note, however, that the age range
allowed for $\varepsilon$\,Indi implies systematic variations of the
model radius. While the younger age (0.8Gyr) increases the radius by
about 3\%, the older age (2Gyr) decreases the radius by 7\% for all
temperatures considered here. Systematic uncertainties up to 15\%
are therefore implicit in the absolute normalization of the model
fluxes due to the age uncertainty of $\varepsilon$\,Indi.
\newline
Burrows, Sudarsy \& Lunine (2003) explore the age and mass
dependance of H$_2$O, CH$_4$ and NH$_3$ molecular bands on MIR
spectra of very cool brown dwarfs with $T_\mathrm{eff} \le 800$\,K.
We have selected a suitable spectral model ($T_\mathrm{eff}=800$\,K,
log $g$ = 5) which represents a brown dwarf with a mass of
$M=25M_{\mathrm{jup}}$ and an age of 1Gyr\footnote{{\tt
zenith.as.arizona.edu/$\sim$burrows/bsl/bsl.html}}. For the hotter
component, the cloud-free model spectrum for an age of 1Gyr
corresponding to $M=46M_{\mathrm{jup}}$ and $M=41M_{\mathrm{jup}}$
for $T_\mathrm{eff}$=1200K and $T_\mathrm{eff}=1100$\,K,
respectively, were supplied by A. Burrows (pers. comm.). Also Allard
et al. (2001) provide model atmosphere spectra in the assumed
parameter range. We concentrate our comparison to their
``cloudless'' models, which include the effects of condensation in
chemical equilibrium, but ignore the effects of dust opacities
altogether. Their model spectra are freely downloadable from the
web\footnote{{\tt
ftp.ens-lyon.fr/pub/users/CRAL/fallard/AMES-Cond-2002}}. Finally,
also SML provide MIR spectra of brown dwarfs.
In fact, their cloudless synthetic spectra for
$T_\mathrm{eff}$=1200K and $T_\mathrm{eff}$=800K (log $g$ = 5) were
combined and compared to the SPITZER spectrum in Roellig et al.
(2004). Both spectra were provided to us by D. Saumon directly.
\newline
For all model spectra we calculated the expected flux densities in
the corresponding VISIR filter passbands by convolving the model
spectrum with the respective filter transmission
profiles\footnote{{\tt http://www.eso.org/instruments/visir/inst/}}.
In Table~\ref{models} we summarize and compare the predicted model
flux densities for the different model atmospheres. In {\bf
boldface} we mark those values that are consistent with our VISIR
measurements (Table~\ref{results}) allowing for a 3$\sigma$ error
range. From the comparison in Table~\ref{models} we infer that most
of the available model spectra for the cooler component Bb are
consistent with the {\sl absolute} MIR photometry. They also
essentially match the {\sl flux ratios} (which are independent of
absolute calibration issues like the radius assumed), but tend to
underestimate the $\mathrm{S{\sc{iv}}\over PAH1}$ color. All models
predict a clear signature of a more or less pronounced NH$_3$
absorption feature, which is also present in our data. The predicted
absolute fluxes appear somewhat ($\approx 20\%$) higher than the
measured ones, but are still within the allowed error range.
\newline
The warmer component Ba is more difficult to reconcile with current
models and assumptions about the $\varepsilon$\,Indi B system. While
cloud-free models, assuming an age of 1Gyr and an effective
temperature of 1200K for Ba as used by Roellig et al. (2004) agree
very well with their SST/IRS spectrum, they appear only {\sl
marginally consistent with our VISIR mid-infrared photometry}. The
spectral shape measured for Ba (decreasing flux with increasing
wavelength) is not reflected in these models. Also, their predicted
MIR fluxes are $\approx 30\% - 100\%$ higher then measured in the
corresponding VISIR passbands. Assuming an age of 2Gyr -- at the
high limit of the allowed age range -- reduces the model radius to
$R=0.087R_\odot$, and the predicted MIR fluxes by 13\%. This is not
sufficient to explain the lower flux densities observed. An age of
5Gyr with a radius of $R=0.082R_\odot$ would be required in order to
match the absolute fluxes.
One straightforward way to reduce the absorption signatures of
NH$_3$ is to increase the temperature of the atmosphere. In fact
Allard's et al. AMES-cond model for $T_\mathrm{eff}=1500$\,K can be
considered to be fully consistent with our {\sl relative}
photometry. However, when we apply the evolutionary models of
Burrows et al. (1997) with a radius of $R=0.09R_\odot$, the {\sl
absolute} photometry is grossly off. $T_\mathrm{eff}=1500$\,K for
component Ba has also been favored by Smith et al. (2003) based on
the analysis of high-resolution near-infrared spectroscopy. This
temperature together with the published luminosity of component Ba
(log $L_\mathrm{bol} = -4.71$, McCaughrean et al. 2004), however,
leads to a radius of only $R=0.062R_\odot$ (Smith et al., 2003).
With this smaller radius, the {\sl absolute} VISIR photometry can
now be made consistent with the spectral model.
But this radius is in contradiction to all known evolutionary models
(see also McCaughrean et al., 2004). Moreover, it also seems
empirically unlikely in view of recent measurements of the
mass-radius relation for very-low mass stars and giant planets,
which prove to be similar (Pont et al., 2005). But more, direct,
determinations of radii in the brown dwarf mass regime are necessary
to completely rule out this possibility.
As can be inferred from Table~\ref{models}, both available
atmosphere models with $T_\mathrm{eff}=1100$\,K tend to agree better
with our measurements, assuming brown dwarf radii derived from
evolutionary models at 1Gyr.
The Burrows et al. models (shown in Figure~\ref{fig2} together with
our photometry) for Ba seem to produce a slightly better match for
the PAH2 measurement because of the presence of a strong absorption
feature (that is less pronounced in the AMES-cond atmosphere).

\begin{table}
      \caption[]{Brown dwarf structural parameters
      according to evolutionary calculations of Burrows et al. (1997) for 1Gyr.}
         \label{parameters}

         \begin{tabular}{lrrcccc}
            \hline
            \hline
            \noalign{\smallskip}
            Object  &  SpT$^\mathrm{a}$ &  $T_\mathrm{eff}$ & log $g$ & log $L_\mathrm{bol}$ & $M$  & $R$ \\
            & &  [K]       & [cm s$^{-2}$]& [$L_\odot$] & [$M_{\mathrm{jup}}$] & [$R_\odot$] \\
            \noalign{\smallskip}
            \hline
            \noalign{\smallskip}
            Ba & T1   & 1200$^\mathrm{b}$ & 5.1 & -4.78    & 46 & 0.093 \\
               &      & 1500$^\mathrm{c}$ & 5.3 & -4.42    & 59 & 0.090$^\mathrm{d}$ \\
               &      & 1100              & 5.1 & -4.92    & 41 & 0.095 \\
            Bb & T6   & 800$^\mathrm{b}$  & 4.8 & -5.42    & 25 & 0.102 \\
           \noalign{\smallskip}
            \hline
         \end{tabular}
         \begin{list}{}{}
        \item[$^{\mathrm{a}}$] Spectral types according to McCaughrean et al. (2003)
        \item[$^{\mathrm{b}}$] $T_\mathrm{eff}$ assumed by Roellig et al. (2004)
        \item[$^{\mathrm{c}}$] $T_\mathrm{eff}$ according to Smith et al. (2003)
       \item[$^{\mathrm{d}}$] Smith et al. (2003) assume
       $R/R_\odot$=0.062 and $L/L_\odot$=-4.71
        \end{list}
   \end{table}

\begin{table}
      \caption[]{MIR flux densities for various model atmospheres.
      Values printed in {\bf boldface} are consistent
      with our observations within $3\sigma$.
      The 800K models refer to $\varepsilon$\,Indi Bb,
      while the warmer models to Ba.}
         \label{models}

         \begin{tabular}{lrrrrrr}
            \hline
            \hline
            \noalign{\smallskip}
            Reference  &  $T_\mathrm{eff}$ & PAH1 & S{\sc{iv}} & PAH2 & $\mathrm{S{\sc{iv}}\over PAH1}$& $\mathrm{S{\sc{iv}} \over PAH2}$ \\
            &  [K]   & [mJy] & [mJy] & [mJy]  & & \\
            \noalign{\smallskip}
            \hline
            \noalign{\smallskip}
            Allard$^{\mathrm{a}}$  & 800  & {\bf 4.32}  & {\bf 3.64}  & {\bf 6.98}  & 0.84 & {\bf 0.52} \\
            Burrows$^{\mathrm{b}}$ & 800  & { 4.76}  & {\bf 4.10}  & {\bf 6.04}  & 0.86 & {\bf 0.68} \\
            Saumon$^{\mathrm{c}}$  & 800  & {\bf 3.68}  & {\bf 2.18}  & {\bf 4.86}  & 0.59 & {\bf 0.45} \\
            \hline
            Allard$^{\mathrm{a}}$   & 1100  & 8.83 & 8.93 & 12.04 & {\bf 1.01} & 0.74 \\
            Burrows$^{\mathrm{d}}$  & 1100  & {\bf 7.96}  & {\bf 8.79} & {\bf 8.83}  & 1.10 & {\bf 1.00} \\
            \hline
            Allard$^{\mathrm{a}}$   & 1200  & 10.67 & 11.36 & 13.21 & 1.06 & 0.86 \\
            Burrows$^{\mathrm{d}}$  & 1200  & 9.13  & 10.36 & 9.99  & 1.13 & {\bf 1.04} \\
            Saumon$^{\mathrm{c}}$   & 1200  & 9.21 & 9.87 & 11.68 & 1.07 & 0.85 \\
            \hline
            Allard$^{\mathrm{a}}$   & 1500  & 17.09 & 16.97 & 16.60 & {\bf 0.99} & {\bf 1.02} \\
            Allard$^{\mathrm{e}}$   & 1500  & {\bf 8.11} & {\bf 8.05} & {\bf 7.88} & {\bf 0.99} & {\bf 1.02} \\

               \noalign{\smallskip}
            \hline
         \end{tabular}
         \begin{list}{}{}
        \item[$^{\mathrm{a}}$] AMES-cond models from Allard et al. (2001)
        \item[$^{\mathrm{b}}$] cloud-free model from Burrows, Sudarsky \& Lunine (2003)
        \item[$^{\mathrm{c}}$] cloud-free models from Saumon, Marley
        \& Lodders (2003, private communication)
        \item[$^{\mathrm{d}}$] cloud-free model from Burrows (private communication)
        \item[$^{\mathrm{e}}$] assuming $R/R_\odot=0.062$
        \end{list}
   \end{table}


\section{Summary}

We report the detection of both components of $\varepsilon$\,Indi B,
the closest brown dwarf binary known, with VISIR at the VLT in three
narrow-band mid-infrared bandpasses located around 8.6$\mu$m,
10.5$\mu$m and 11.3$\mu$m. We are able to determine accurate
mid-infrared absolute photometry for both components independently,
with an error level of 5-10\% for the brighter component Ba, and
10-20\% for the fainter Bb. Relative photometry and flux ratios can
be measured with even higher accuracy. Our data show that component
Bb has a prominent absorption feature around 10.5$\mu$m, most likely
explained by NH$_3$. We then compare our MIR photometry with
atmospheric model spectra, using the well-known distance,
metallicity and age of the $\varepsilon$\,Indi B system as main
input parameters. The MIR emission of the cool component Bb appears
to be fully consistent with current atmosphere models assuming an
effective temperature of $T_\mathrm{eff}=800$\,K. The warmer
component Ba appears only marginally consistent with
$T_\mathrm{eff}=1200$\,K, a temperature that has been inferred from
SPITZER spectroscopic observations of the combined
$\varepsilon$\,Indi B system (Roellig et al., 2004), and from
near-infrared photometry (McCaughrean et al. 2004), if we assume a
canonical age of 1Gyr for the $\varepsilon$\,Indi B system. We
instead favor a slightly lower effective temperature of
$T_\mathrm{eff}=1100$\,K to reconcile the absolute MIR fluxes and
the spectral shape, which does not show any evidence for NH$_3$
absorption. A higher effective temperature of
$T_\mathrm{eff}=1500$\,K which would agree with the spectroscopic
temperature derived for the $\varepsilon$\,Indi B system by Smith et
al. (2003), implies a non-physical small radius of
$\varepsilon$\,Indi Ba, and is therefore unlikely.


\begin{acknowledgements}
We like to thank the VISIR science verification team and the ESO
director general for allocating observing time to this project. We
appreciate to use the model spectra of F. Allard, A. Burrows and D.
Saumon in electronic form. A. Burrows to supplied his 1100K and
1200K spectrum, and D. Saumon his 800K and 1200K model directly. The
referee, A. Burgasser, helped to improve the paper.

\end{acknowledgements}


\end{document}